\documentclass[twocolumn]{aastex62}
\usepackage{amsmath,amstext}
\usepackage{units}
\usepackage{url}
\usepackage{xspace}
\usepackage{graphicx}

\newcommand{\E}[1]{\times10^{#1}}
\newcommand{\msol}{ \, M_\odot}

\newcommand{\bi}{\begin{itemize}}
\newcommand{\ei}{\end{itemize}}
\newcommand{\commentOut}[1]{}

\newcommand{\mesa}{\texttt{MESA}\xspace}

\shortauthors{Shen}

\begin{document}


\title{\bf \Large{The Evolution of Hypervelocity Supernova Survivors and the Outcomes of Interacting Double White Dwarf Binaries}}

\correspondingauthor{Ken~J.~Shen}
\email{kenshen@astro.berkeley.edu}

\author[0000-0002-9632-6106]{Ken~J.~Shen}
\affiliation{Department of Astronomy and Theoretical Astrophysics Center, University of California, Berkeley, CA 94720, USA}


\begin{abstract}

The recent prediction and discovery of hypervelocity supernova survivors has provided strong evidence that the ``dynamically driven double-degenerate double-detonation'' (D$^6$) Type Ia supernova scenario occurs in Nature.  In this model, the accretion stream from the secondary white dwarf in a double white dwarf binary strikes the primary white dwarf violently enough to trigger a helium shell detonation, which in turn triggers a carbon/oxygen core detonation.  If the secondary white dwarf survives the primary's explosion, it will be flung away as a hypervelocity star.  While previous work has shown that the hotter observed D$^6$ stars can be broadly understood as secondaries whose outer layers have been heated by their primaries' explosions, the properties of the cooler D$^6$ stars have proven difficult to reproduce.  In this paper, we show that the cool D$^6$ stars can be explained by the Kelvin-Helmholtz contraction of helium or carbon/oxygen white dwarfs that underwent significant mass loss and core heating prior to and during the explosion of their  white dwarf companions.  We find that the current population of known D$^6$ candidates is consistent with $\sim2\%$ of Type Ia supernovae leaving behind a hypervelocity surviving companion.  We also calculate the evolution of hot, low-mass oxygen/neon stars and find reasonable agreement with the properties of the LP 40-365 class of hypervelocity survivors, suggesting that these stars are the kicked remnants of  near-Chandrasekhar-mass oxygen/neon white dwarfs that were partially disrupted by  oxygen deflagrations.  We use  these results  as motivation for  schematic diagrams showing speculative outcomes of interacting double white dwarf binaries, including long-lived merger remnants, Type Ia supernovae, and several kinds of peculiar transients.

\end{abstract}


\section{Introduction and motivation}

There is broad consensus that Type Ia supernovae (SNe~Ia) occur via the thermonuclear explosion of a C/O white dwarf (WD) in a stellar system with multiple components; for recent reviews, see \cite{liu23b} and \cite{ruit24a}.  However, debate has continued for half a century regarding the identity of the companion star(s) (e.g., H- or He-rich non-degenerate donors or He- or C/O-core WDs), how and whether mass is transferred (e.g., stably over secular timescales; unstably in the lead-up to a merger; or negligibly in a stellar collision), and how the explosion unfolds (e.g., via a core deflagration that transitions into a detonation; a He shell detonation followed by a C/O core detonation, also known as a double detonation; or a direct C detonation).

The past decade has seen the emergence of the ``dynamically driven double-degenerate double-detonation'' (D$^6$) scenario in which dynamical-timescale mass transfer from a secondary WD ignites a double detonation in a primary WD \citep{guil10,dan11,dan12,rask12,pakm13a,pakm22a,moll14a,tani18b,tani19a,raja25a}.\footnote{Here and throughout this paper, we refer to the less massive WD as the secondary and the more massive WD as the primary.} In addition to yielding better and better matches to observations as simulations become more accurate \citep{fink10,krom10,sim10,shen18a,poli19a,town19a,gron20a,gron21a,boos21a,shen21a,shen21b,boos24b,coll24a}, this model has strong support in the prediction and discovery of hypervelocity stars that are best interpreted as the surviving companions of double WD systems in which the primary WD exploded as a SN~Ia, slingshotting the secondary WD at velocities $>1000 {\rm \, km \, s^{-1}}$ \citep{shen18b,elba23a}.  Several of these candidate D$^6$ survivors can be adequately modeled as former companion WDs whose surface layers have been shock-heated by their primary WD's explosions \citep{bhat25a}, but the coolest survivors have resisted a convincing, quantitative explanation.

Complicating matters further, it has become increasingly clear that the secondary WD may undergo its own double detonation due to the impact from the primary WD's ejecta \citep{tani19a,pakm22a,boos24a}, in which case the secondary would not survive as a hypervelocity star.  \cite{shen24a} found that nearly all C/O WDs are born with enough He on their surfaces to sustain a He shell detonation, and that these He detonations should lead to a C/O core detonation \citep{shen14a,ghos22a}.  Thus, the shock from the primary's explosion should be capable of destroying most secondaries if the impact is strong enough.  

The only exceptions leading to a hypervelocity surviving companion would be cases in which the secondary does not possess a massive-enough He layer to sustain a detonation at the time of the primary's explosion.  There are three ways this outcome might come to pass: if the secondary is $\gtrsim 1.0 \msol$ and is not born with enough He on its surface \citep{shen24a}; if it is a He-core WD;\footnote{The He-core WDs in several studies \citep{papi15a,tani19a,boos24a}  detonated due to the impact from their primary WD's explosions but only for artificially reduced separations between the two WDs.  Further studies are needed to fully explore this possibility.} or if the secondary loses significant mass, including any He-rich surface layers, prior to the WD's explosion.

The latter scenario may occur if the mass ratio of the double WD binary is relatively low, because the accretion stream from the secondary will strike the primary at a shallow angle, resulting in lower temperatures and densities at the onset of mass transfer.  The ignition of the primary's He shell detonation may thus require the merging process to proceed further; as mass transfer continues and the donor loses more mass, the accretion rate and ram pressure increase, making a detonation more likely \citep{dan11}.

The loss of copious mass affects the secondary in several ways.  It will expand beyond its Roche lobe and subtend a larger solid angle with respect to the primary WD's ejecta, and thus it will capture a greater fraction of the ejecta's shock energy.  Moreover, its central pressure will  be decreased, so the ram pressure from the ejecta shock can significantly heat the survivor to a greater depth.  Finally, its lower gravitational potential enables more of its mass to be stripped off by the shock's impact.  The combination of all of these effects allows for a scenario in which a surviving companion WD is converted into a much lower-mass star that is significantly heated all the way to its center. 

Motivated by this possibility, we calculate the evolution of centrally heated, fully convective stars in this paper and compare their observational parameters to the known hypervelocity D$^6$ survivors.  We also construct similar models but with O/Ne-rich compositions to compare to the LP 40-365 (GD 492) hypervelocity stars \citep{venn17a,radd18a,radd18b,radd19a,elba23a}.  These stars are hypothesized to have been the surviving remnants of near-Chandrasekhar-mass WDs that underwent partial deflagrations, possibly leading to Type Iax supernovae.  The star that remains after the SN ejecta has dissipated may have lost much of its mass and been significantly heated, and thus might also be modeled as an initially fully convective star.

We describe our models and simulations in Section \ref{sec:mesa}. In Section \ref{sec:rates}, we use the results of these calculations  to refine previous estimates of the rate of SNe~Ia that leave behind hypervelocity survivors.  In Section \ref{sec:mvsm}, we qualitatively delineate the regions of parameter space that may lead to these and other interacting double WD outcomes, and we conclude in Section \ref{sec:conc}.


\section{\texttt{MESA} simulations and results}
\label{sec:mesa}

We perform our stellar evolutionary calculations using \mesa\citep{paxt11,paxt13,paxt15a,paxt18a,paxt19a}\footnote{https://docs.mesastar.org, version 10398.}  and custom opacity tables.  \mesa includes opacity tables for He-rich and C/O-rich matter, but to account for lower temperatures and for O/Ne-rich material, we calculate our own opacities (for all temperatures and compositions for consistency).  Following the procedure outlined in \cite{schw16a}, we use atomic data compiled in \texttt{CMFGEN} \citep{hill11a} to calculate Rosseland mean opacities including bound-bound, bound-free, free-free, and electron scattering processes, assuming local thermodynamic equilibrium.

We consider three different compositions: He-rich and C/O-rich, as appropriate for the inner regions of He and C/O WDs, respectively, and an average of the O/Ne-rich compositions derived by \cite{radd19a} for the atmospheres of three LP 40 stars.  The He-rich composition is taken from the center of a $1.0 \msol$ star that has been evolved with \mesa until the formation of a degenerate He core.  The initial composition is taken to be the bulk solar composition from \cite{aspl09a} with photospheric metal ratios.  The C/O-rich composition is similarly calculated by evolving a $3.0 \msol$ star to the point of He exhaustion in the core.  The mass fractions for these three compositions are listed in Table \ref{tab:comp}.

\begin{deluxetable*}{ccccccccccc}
\tablecolumns{11}
\tablecaption{Compositions for opacities and evolutionary calculations used in this work.  Only elements with mass fractions $>10^{-3}$ are considered.\label{tab:comp}}
\tablehead{
\colhead{Description} & \colhead{$X_{\rm He}$} & \colhead{$X_{\rm C}$} & \colhead{$X_{\rm N}$} & \colhead{$X_{\rm O}$} & \colhead{$X_{\rm Ne}$}  & \colhead{$X_{\rm Na}$}  &\colhead{$X_{\rm Mg}$} &\colhead{$X_{\rm Al}$}  &\colhead{$X_{\rm Si}$} &\colhead{$X_{\rm Fe}$}}
\startdata
He-rich & 0.987  & ---  & $7.08\E{-3}$ & $2.72\E{-3}$ & $1.41\E{-3}$ & --- & ---  & --- & --- & $1.41\E{-3}$  \\
C/O-rich & ---  & 0.354  & --- & 0.629 & 0.0154 & --- & ---  & --- & --- & $1.41\E{-3}$  \\
O/Ne-rich & ---  & ---  & --- & 0.306 & 0.631 & $9.40\E{-3}$ & 0.0475  & $3.09\E{-3}$ & $1.55\E{-3}$ & $1.52\E{-3}$  \\
\enddata
\end{deluxetable*}

With these opacities in hand, we construct hot, fully convective stars of various masses with the same compositions and evolve them with \mesa.   The input files to recreate our results are publicly available at \dataset[doi:10.5281/zenodo.14814815]{https://doi.org/10.5281/zenodo.14814815}. We heat the initial models until the outgoing luminosity is at least $10 \, L_\odot$; the precise initial entropy and corresponding luminosity of the models are unimportant as long as the models have an initial Kelvin-Helmholtz timescale shorter than the ages of interest, which will indeed be the case as shown in the next sections.  Our models are similar in spirit to the highest envelope fraction models calculated by \cite{zhan19a}, but we input only enough energy to expand our initial models and not enough to lead to pathological outcomes.


\subsection{Model comparisons to D$^6$ survivors}
\label{sec:d6}

\begin{figure}
  \centering
  \includegraphics[width=\columnwidth]{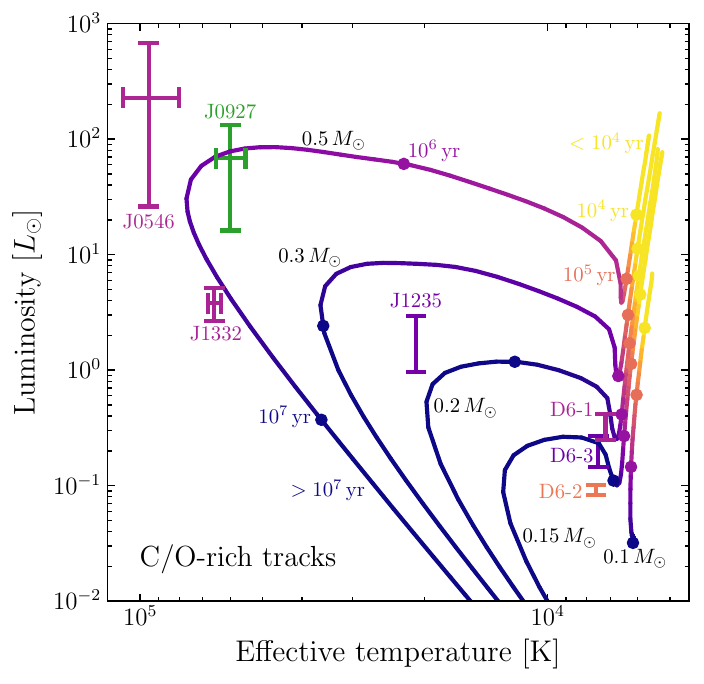}
	\caption{Luminosity vs.\ effective temperature of hypervelocity stars with compositions similar to D6-1, D6-2, and D6-3, shown as error bars, and evolutionary tracks of initially fully convective C/O-rich stars from \mesa, with masses as labeled.  Colors and circles correspond to the labeled ages.  J0927-6335 is displayed in green because it has a negative travel time from the midplane.}
	\label{fig:traj_CO}
\end{figure}

Figure \ref{fig:traj_CO} shows the resulting evolutionary tracks of our C/O-rich models in the Hertzsprung-Russell diagram, with masses ranging from $0.1$ to $0.5 \msol$, as labeled.  Colors correspond to ages; labeled circles demarcate specific times.    Also shown are most of the known D$^6$ candidate survivors.  Effective temperatures for the three hottest D$^6$ stars are taken from \cite{wern24a}.\footnote{We note that the H-rich atmosphere of J1332-3541 \citep{wern24a} can be explained as arising from  a tiny amount ($> 10^{-16} \msol$; S.\ Blouin 2024, private communication) of Bondi-Hoyle-Lyttleton accretion from the interstellar medium, which is especially plausible if it has passed through the Galactic midplane.}  We use their derived luminosities, but scaled by the square of the ratio of their median SED distances (which are derived from  evolutionary tracks that may not correspond to these peculiar stars) to the median \emph{Gaia} distances.  Colors correspond to travel times from the midplane, $t_{\rm midplane}$, as given in \cite{elba23a}.  J0927-6335 is shown as a green error bar because its travel time is negative (i.e., it is moving towards the plane).  The luminosity and effective temperature for J1235-3752 are taken from \cite{elba23a}; no error bar is given for the temperature.

The properties of the three coolest D$^6$ stars from \cite{shen18b} are estimated using the dependence of effective temperature and bolometric correction on $G_{\rm BP}-G_{\rm RP}$ colors for He atmosphere WDs in \cite{gent21a}'s catalog.  These stars are clearly not He atmosphere WDs, so the displayed properties are only approximate.  No extinction corrections are made, and only errors derived from parallax uncertainties are shown for these three stars.  Again, $t_{\rm midplane}$ is used as a proxy for age, except for D6-2, whose age is given by its travel time from the SN remnant G70.0–21.5 \citep{fese15a,shen18b}.  J1637+3631, discovered by \cite{radd19a} and likely another D$^6$ star (Hollands et al., in preparation), is not shown as it does not have an accurate luminosity estimate.

Figure \ref{fig:traj_CO} shows that convective C/O stars evolve down the C/O version of the \citet{haya61a} track at a nearly constant temperature of $\simeq 6000$\,K.  When their cores become radiative, the stars execute a bounce towards hotter temperatures and  higher luminosities, contracting until they become degenerate and evolve along the WD cooling sequence.  The $0.1 \msol$ model encounters numerical issues shortly after an age of $10^7$\,yr and thus is not followed further in time.

The luminosities and effective temperatures for D6-1 and D6-3 are reasonably well produced at their respective ages by the $0.15 \msol$ track.  The agreement with J1235-3752 is somewhat worse; a model with a mass $\simeq 0.25 \msol$ would evolve through its position, but at a later age.

The parameters of the three hottest D$^6$ stars can be matched by a model with a mass $\gtrsim 0.5 \msol$, possibly at the correct age; however, it is not obvious that such a relatively high-mass, initially fully convective star can be created.  While it is plausible that a  companion WD could lose nearly all of its mass and end up as a very low-mass $\sim 0.2 \msol$ object, rendering it very diffuse and subject to significant heating by the primary WD's ejecta impact, it is less obvious that a $0.5 \msol$ survivor could be converted into a fully convective star because its central pressure is much larger (although future hydrodynamic simulations should be conducted to better quantify these outcomes).  These hottest survivors may instead come from systems in which the companion WD is nearly unperturbed at the time of the primary's explosion, but does not possess enough of a He shell to undergo its own detonation.  In this case, only the surface layers will be significantly heated by the impact, and the further evolution will be similar to the models calculated by \cite{bhat25a}.

The young age of D6-2 also presents a problem for these C/O-rich models.  The Kelvin-Helmholtz timescale of a star undergoing gravitational contraction depends on the  mass, $M$, effective temperature, $T_{\rm eff}$, and luminosity, $L$, as
\begin{align}
t_{\rm K-H} \propto \frac{ M^2 T_{\rm eff}^2}{L^{3/2}} .
\end{align}
D6-2's effective temperature is roughly the same as for D6-1 and D6-3, and its lower luminosity and age yield a mass that is $\sim 5$ times lower: assuming a mass of $0.15 \msol$ for D6-1 and D6-3 implies a mass of just $0.03 \msol$ for D6-2.  While we were unable to produce a $0.03 \msol$ C/O-rich model in \mesa for numerical reasons, it is clear that  the Hayashi track for this small mass would be much redder and colder than D6-2's effective temperature.

\begin{figure}
  \centering
  \includegraphics[width=\columnwidth]{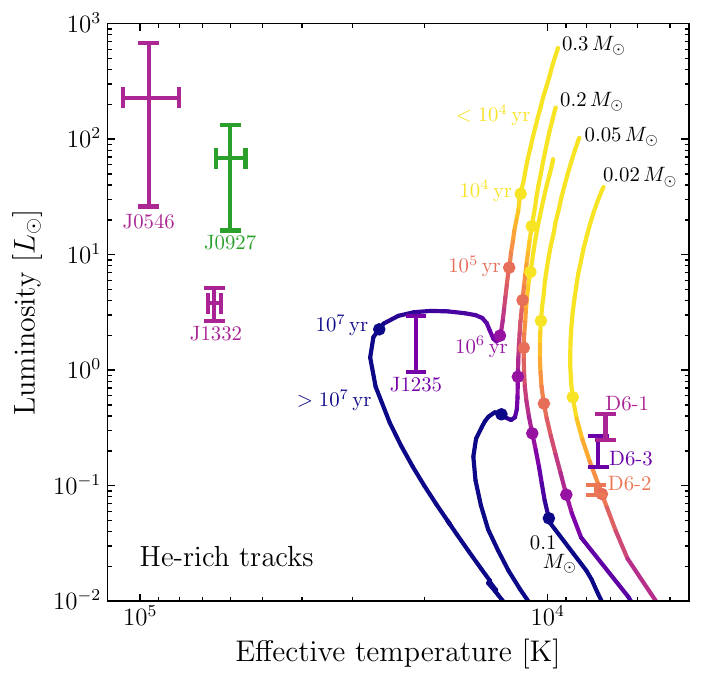}
	\caption{Same as Fig.\ \ref{fig:traj_CO} but with He-rich \mesa tracks of different masses.}
	\label{fig:traj_HeN}
\end{figure}

Figure \ref{fig:traj_HeN} shows the results of our He-rich simulations, again compared to known D$^6$ survivors.  The hottest D$^6$ stars are out of reach of even the highest mass case we consider, $0.3 \msol$, as He WDs have a maximum mass of $0.45 \msol$, and the companion WD must lose significant mass to be converted into a fully convective star by the primary's ejecta impact.  While J1235-3752 might be plausibly explained as a $0.3 \msol$ He-rich star based on its Hertzsprung-Russell diagram position and age, its spectrum shows no He features \citep{elba23a}, which should be apparent at this effective temperature if its atmosphere were He-rich.  D6-1 and D6-3 are also not well-explained with a He-rich composition, since their ages do not match the low-mass Hayashi track to which they correspond.

However, D6-2's effective temperature, luminosity, and age agree well with the cooling evolution of a very low-mass $0.02 \msol$ He-rich object.   While He is not directly seen in D6-2's spectrum, its relatively low temperature $<10000$\,K precludes the formation of He features.  \cite{chan22a} analyzed D6-2's spectrum and concluded that the number density ratio of C/He could not even be as high as $10^{-5}$ due to the formation of unobserved molecular C$_2$ bands.  This is in contrast to observed atomic C  features, leading the authors to suggest the presence of circumstellar gas in which the atomic C  features arise.   However, if D6-2's mass is indeed as low as $0.02 \msol$, its surface gravity and photospheric pressure and density are much lower than assumed by \cite{chan22a}, and a number density ratio of C/He of $1.4\E{-5}$, as appropriate for the center of a He WD, may avoid the formation of molecules while still giving rise to the observed atomic C spectral features.  Future detailed analysis of D6-2's spectrum, and of all of the D$^6$ stars' spectra, will shed light on these possibilities.

The low masses implied for D6-1 and D6-3 (and possibly J1235-3752) and the very low mass derived for D6-2 require these former companion WDs to have lost nearly all of their mass between the onset of mass transfer and the aftermath of their primary WD's explosion.  This would make sense if the ignition of the primary WD required the merging process to go nearly to completion.  

One possibility is if  the primary WD is $\gtrsim 1.0 \msol$ and does not possess a dense enough He layer to allow for a shell detonation \citep{shen24a}.  In this case, a core detonation might still occur via a direct C ignition \citep{pakm10,pakm11,pakm12b}. However, this would require a direct impact accretion stream with a much higher ram pressure to reach the densities and temperatures necessary to ignite C.

As discussed by \cite{dan11}, the accretion rate and ram pressure of the accretion stream during a double WD merger increase up to the endpoint of the companion WD's complete tidal disruption.  Thus, a direct C ignition might be achieved, but it would require the mass accretion phase to continue for longer, and  the companion would therefore lose more mass prior to the explosion than if the primary's He shell were able to be detonated.  \cite{dan11} also note that the companion reaches higher velocities as it approaches the final plunge, up to the primary WD's escape velocity.  This may explain the large velocities inferred for D6-1 and D6-3 (and possibly J1235-3752 and J0927-6335), which otherwise push the upper limit of double WD orbital velocities \citep{baue21a,elba23a}.

The large amount of mass lost by the companion implies that it avoids undergoing its own explosion because it has lost its He layer and instead allows it to be converted into a fully convective star.  One reason for this is that  mass transfer has progressed closer to the point of complete tidal disruption, meaning  the deposition of a larger amount of energy from tidal dissipation.  Furthermore, the expansion of the companion due to mass loss allows it to intercept a larger cross section of the primary WD's SN ejecta and correspondingly a larger amount of shock energy.  Perhaps most importantly, the companion's central pressure following mass loss is much lower than for an unperturbed WD.  Thus, the ram pressure from the SN ejecta's impact can significantly alter the internal conditions of the surviving companion.  The combination of these effects may lead to the transformation of the companion WD into a much lower-mass, fully convective hypervelocity runaway star.\footnote{If this scenario is indeed responsible for producing some of the D$^6$ stars, their designations are a misnomer because neither WD undergoes a double detonation.}

The large amount of companion mass lost in systems in which the primary undergoes a direct C ignition will also have important effects on the resulting SN.  Analogous work has studied thermonuclear explosions inside massive C/O-rich envelopes \citep{rask14a,noeb16a,fitz23a}, with the difference being here we consider the possibility of  a low-mass surviving companion remnant.  The excess material surrounding the exploding primary will slow the ejecta and convert kinetic energy into thermal energy, making the light curves brighter and more slowly evolving.  The spectra will also be affected, with lower velocities and signatures of unburned C and O.

Such overluminous, partially interaction-powered SNe~Ia with unburned C/O material match the characteristics of SN~2003fg-like events \citep{howe06a}.  Thus, we speculate that interacting double WDs with high-mass C/O primaries $ \gtrsim 1.0 \msol$ lead to SN~2003fg-like SNe if a direct C  detonation can be ignited prior to the complete disruption of the companion.  We note that this scenario also aligns with the interesting case of SN~2020hvf, a SN~2003fg-like transient that showed evidence for a late-phase wind from a surviving companion \citep{sieb23a}.

Assuming D6-2 is indeed He-rich, its initial mass is constrained to be $ \leq 0.45 \msol$ if it was  a WD at the onset of mass transfer.  The accretion stream from a low-mass He WD impacts the primary at a more grazing angle than the stream from a a more massive C/O secondary WD, leading to shallower deposition of the ram pressure.  Thus, unlike for the C/O-rich survivors, the companion WD that gave rise to D6-2 may not have been able to initiate a direct C  ignition in a high-mass primary WD.  Instead, D6-2 may have lost significant mass because the merging process needed to become violent enough merely to initiate a He shell detonation in the primary WD.  The impact from the primary's ejecta then stripped further mass, leaving the very low-mass D6-2 as a hypervelocity survivor.

However, its relatively low ejection velocity of $1050 \, {\rm km \, s^{-1}}$ presents a puzzle.  The shock from the SN ejecta must have deposited   momentum and should naively have accelerated D6-2's low mass of  $0.02 \msol$ to higher velocities.  Future hydrodynamic simulations will quantify this effect, as well as accounting for the non-instantaneous passage of the SN ejecta \citep{brau24a}.

An alternate scenario to explain D6-2 is if it was a non-degenerate He star that donated $\sim 0.1 \msol$ to a WD, eventually triggering the WD's explosion via convective shell-burning that transitions into a double detonation \citep{nomo82a,iben87b,it91,baue19a,wong24a}. If the donor's mass at the time of the explosion  was whittled down to $0.1 \msol$, it is possible that the impact from the SN ejecta could strip off the majority of the remaining mass and leave a very low mass $\sim 0.02 \msol$ survivor.  However,  D6-2's low observed ejection velocity still remains an issue, since the impact from the SN ejecta should give such a low mass remnant a large kick.  Future spectral modeling will possibly yield a clue, as the C abundance at the center of a He-burning star is higher than within a He WD.

We note that the problem with D6-2's relatively low velocity would be alleviated if D6-2's mass were not inferred to be so small from its very short lifetime, which is due to its association with the SN remnant G70.0–21.5.  There is a possibility that this association is a coincidence, and that D6-2's actual age is much longer than $10^5$\,yr, which would allow its Kelvin-Helmholtz lifetime to be longer and its mass to be larger, thereby enabling it to avoid such a large kick from the ejecta impact.  However, the fact that D6-2 passed near the three-dimensional position of  G70.0-21.5 at the time of its explosion and the low rate of SNe at G70.0-21.5's height of $\sim 400 \,$pc from the Galactic midplane suggest that a chance alignment is improbable.  Furthermore, a longer travel time would put D6-2's birthplace at an even larger distance from the Galactic midplane, which is unlikely due to the exponential decrease in stellar density.  Still, a better estimate of the probability of a random coincidence  should be quantified in future work.

The scenario outlined above -- namely, that D6-2 was a He companion WD that lost much of its mass prior to igniting a He shell detonation in its primary WD -- may also be applicable to low-mass C/O companion WDs, since they, too, will have more grazing direct impact accretion streams than for higher-mass companions.  Thus, there may be very low-mass survivors analogous to D6-2 but with C/O-rich compositions.  Comparing Figures \ref{fig:traj_CO} and \ref{fig:traj_HeN}, we see that a C/O-rich survivor with a similar mass to D6-2 would evolve down an even colder Hayashi track, perhaps redder than the color constraint used by \cite{elba23a}.  We thus encourage future efforts aimed at discovering hypervelocity survivors with \emph{Gaia} to expand the search parameter space to redder objects (although this does introduce more false positives).


\subsection{Model comparisons to LP 40-365 survivors}

\begin{figure}
  \centering
  \includegraphics[width=\columnwidth]{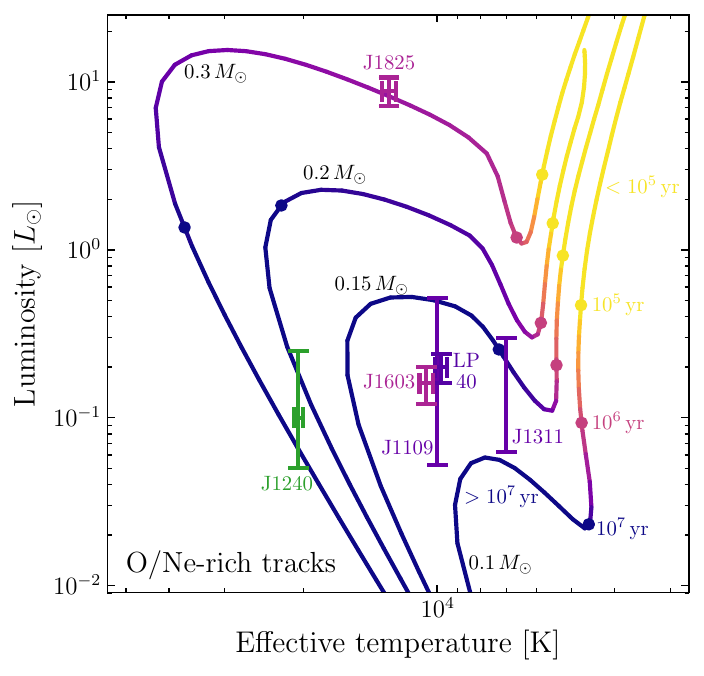}
	\caption{Same as Fig.\ \ref{fig:traj_CO} but showing LP 40-365 stars and  O/Ne-rich \mesa tracks of different masses.  Note that the color - age relationship is also different.}
	\label{fig:traj_ONeMg}
\end{figure}

In Figure \ref{fig:traj_ONeMg}, we show evolutionary tracks for our O/Ne-rich models.  Note that colors correspond to different ages than in Figures \ref{fig:traj_CO} and \ref{fig:traj_HeN}.  Error bars show the properties of the LP 40-365 stars, taken from \cite{radd19a} and \cite{elba23a}.  We do not display J0905+2510, as its distance and thus luminosity are not well-constrained.  The luminosity of J1240+6710 \citep{kepl16a,gaen20a}, a possible member of the LP 40-365 class, is derived using the same procedure as for the three coolest D$^6$ stars described in the previous section.  Its age is unknown because it is bound to the Galaxy and may have crossed the midplane multiple times in the past, so it is displayed in green.

We see that our models with masses between $0.1$ and $0.3 \msol$ can explain the luminosities and effective temperatures of the LP 40-365 stars.  However, with the exception of J1825-3757, the ages do not match well.  The cluster of observed survivors with temperatures $7000-11000$\,K (not including J1240+6710) have midplane ages of $2-5$\,Myr, while the $0.1$ and $0.15 \msol$ tracks do bracket this grouping but at an age $>10$\,Myr.  Still, the low masses for all of these tracks agree with those derived from spectral modeling by \cite{radd19a} and \cite{gaen20a} and are much lower than the minimum initial  O/Ne WD mass $\gtrsim 1.1 \msol$ \citep{mura68a,dohe15a}.  These low masses corroborate a scenario in which a degenerate O/Ne core approached the Chandrasekhar mass and underwent a partial O deflagration that unbound most of the star but left a small, kicked remnant.


\section{The fraction of Type Ia supernovae that leave behind a surviving companion}
\label{sec:rates}

The evolutionary tracks calculated in the previous section allow for a refined estimate of the rate of SNe~Ia that leave behind a detectable surviving companion \citep{shen18b,elba23a}. We model the Milky Way as a disk with a scale height of 300\,pc, a radial scale length of 2.6\,kpc, a total stellar mass of $5\E{10} \msol$ \citep{blan16a}, and a SN~Ia rate of $5\E{-3}$\,yr$^{-1}$.  We consider two populations of D$^6$ survivors: one made up of survivors similar to D6-2, with  ejection velocities of $1050 \, {\rm km \, s^{-1}}$ and luminosity evolutions given by the history of the $0.02 \msol$ He-rich track, and another population of survivors with velocities of $2300 \, {\rm km \, s^{-1}}$ and luminosities given by the $0.15 \msol$ C/O-rich track, as appropriate for D6-1, D6-3, and possibly J1235-3752.  Following \cite{elba23a}, we assume an extinction in the Gaia bandpass of $A_G = 0.85 \, {\rm mag \, kpc^{-1}}$ for stars within $10^\circ$ of the Galactic plane and find the number of expected survivors at the present day with Gaia magnitude $M_G<20$\,mag, proper motion $\mu>50 {\rm \, mas \, yr^{-1}}$, and tangential velocity $v_{\rm tan}>600 \, {\rm km \, s^{-1}}$, assuming every SN~Ia yields a surviving companion.

\begin{figure}
  \centering
  \includegraphics[width=\columnwidth]{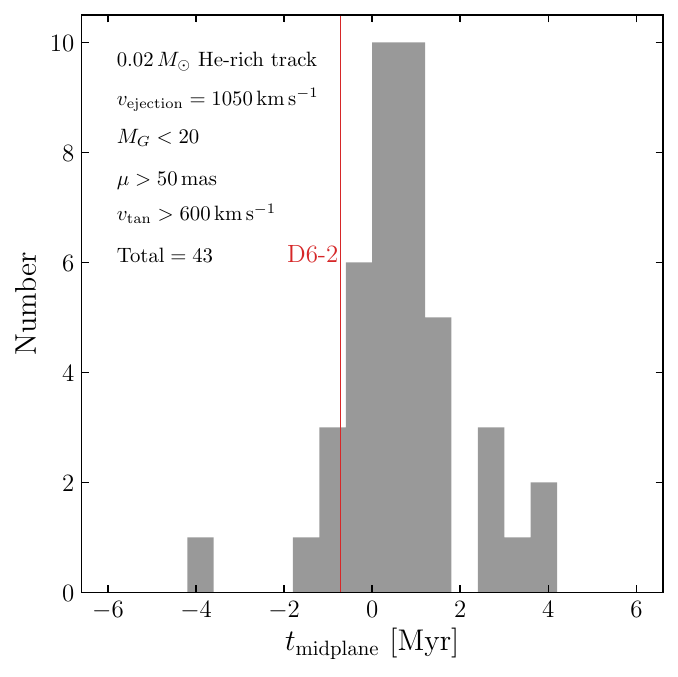}
	\caption{Histogram of detectable hypervelocity survivors, assuming all SNe~Ia eject a $0.02 \msol$ He-rich companion at a velocity of $1050 \, {\rm km \, s^{-1}}$.  The criteria for detection are as listed in the figure and described in the text.  D6-2's midplane time is as labeled.}
	\label{fig:hist02}
\end{figure}

\begin{figure}
  \centering
  \includegraphics[width=\columnwidth]{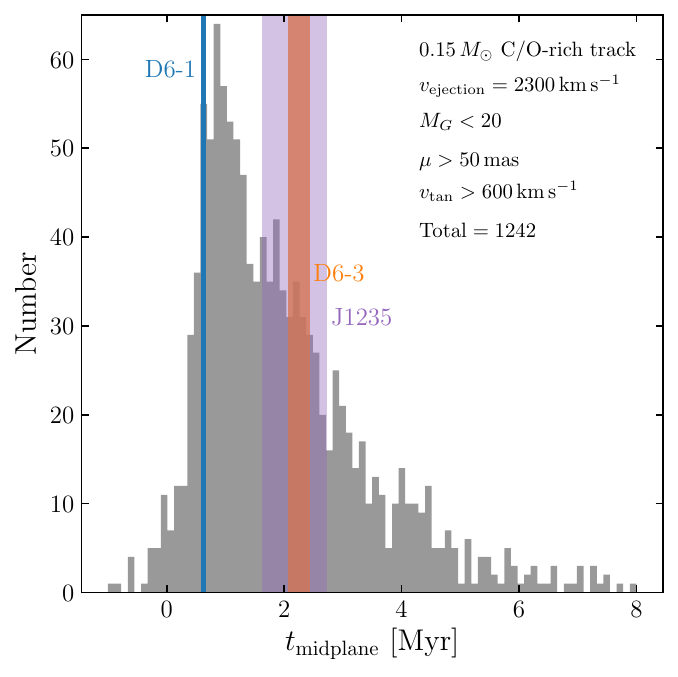}
	\caption{Same as Fig.\ \ref{fig:hist02}, but for $0.15 \msol $ C/O-rich survivors ejected at $2300 \, {\rm km \, s^{-1}}$.  The $1-\sigma$ range of midplane times of D6-1, D6-3, and J1235-3752 are as labeled.}
	\label{fig:hist15}
\end{figure}

The resulting histograms binned by apparent travel time from the midplane, $t_{\rm midplane}$, are shown in Figures \ref{fig:hist02} and \ref{fig:hist15} for the $0.02 \msol$ He-rich  track and the $0.15 \msol$ C/O-rich  track, respectively.  The distributions are subject to small number statistics; representative populations are chosen that match the mean number of survivors for each of the two cases.  Note that midplane times can be negative if a survivor is ejected towards the plane and is observed before it has passed through it, as is the case for D6-2.

If every SN~Ia yielded a $0.02 \msol$ He-rich survivor, Figure \ref{fig:hist02} shows that we would expect to have detected $\sim 40$ stars like D6-2.  Our discovery of just one such star implies that a He-rich companion survives $\sim 2\%$ of SNe~Ia.  D6-2's negative midplane age is somewhat uncommon but not completely atypical: $\sim 25\%$ of the samples in the simulated population have a negative midplane age.

From Figure \ref{fig:hist15}, we see that the three stars that are likely low-mass C/O-rich survivors are consistent with SNe~Ia ejecting such stars only $0.2\%$ of the time.  As previously discussed  (and see \citealt{shen24a}), these survivors may have triggered SN~2003fg-like SNe Ia, losing much of their mass in the process.  Within small number statistics and our model uncertainties, the discovery of three stars like these is broadly compatible with the rate of SN~2003fg-like SNe, which make up $0.8\% \pm 0.2\%$ of SNe~Ia \citep{dimi24a}. 

A similar calculation can be performed for the LP 40-365 survivors, but since these stars move at lower velocities, the straight-line approximation we use for the D$^6$ survivors is less applicable; indeed, J1825-3757 and J1240+6710, which may be a member of the LP 40-365 class, are still bound to the Milky Way \citep{kepl16a,radd19a,gaen20a}.  \cite{radd19a} performed a rate calculation  including the effect of the Galaxy's gravitational potential and using evolutionary tracks from \cite{zhan19a} and estimated that there should be $\sim 60$ LP 40-365 survivors with $M_G<18$.  Similar calculations using the evolutionary tracks we have developed here and accounting for the Galaxy's potential as well as the birth velocity offset due to the rotation of stars in the disk will be the subject of future work.


\section{Outcomes of interacting double WD binaries}
\label{sec:mvsm}

\begin{figure}
  \centering
  \includegraphics[width=\columnwidth]{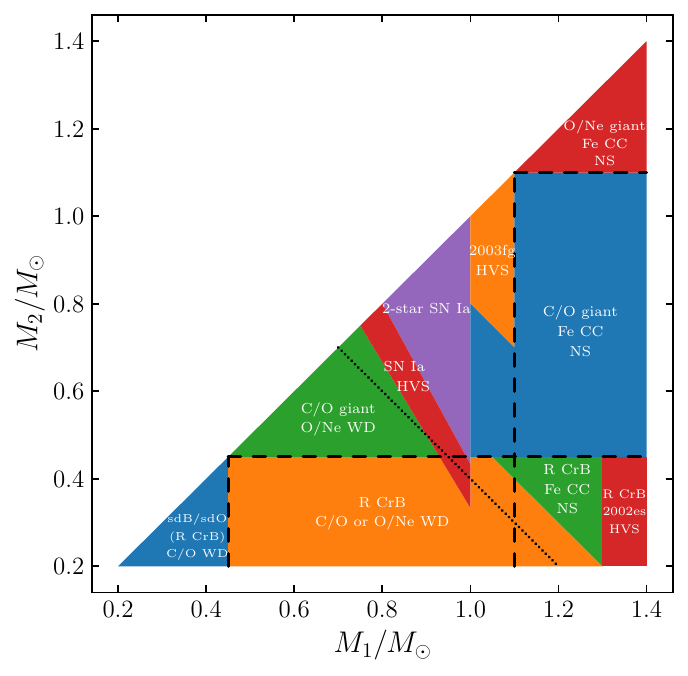}
	\caption{Speculative outcomes of interacting double WDs binaries, as a function of primary WD mass, $M_1$, and secondary WD mass, $M_2$.  All binaries are assumed to undergo dynamical timescale mass transfer \citep{shen15a}, and the primary WD is assumed to possess its birth structure at the time of interaction \citep{shen24a}.  Merger remnants are assumed to lose $\sim 10\%$ of their mass during their evolution. Dashed lines delineate regions of different compositions, and the dotted line demarcates a total combined mass of $1.4 \msol$.}
	\label{fig:mvsm_noaddhe}
\end{figure}

\begin{figure}
  \centering
  \includegraphics[width=\columnwidth]{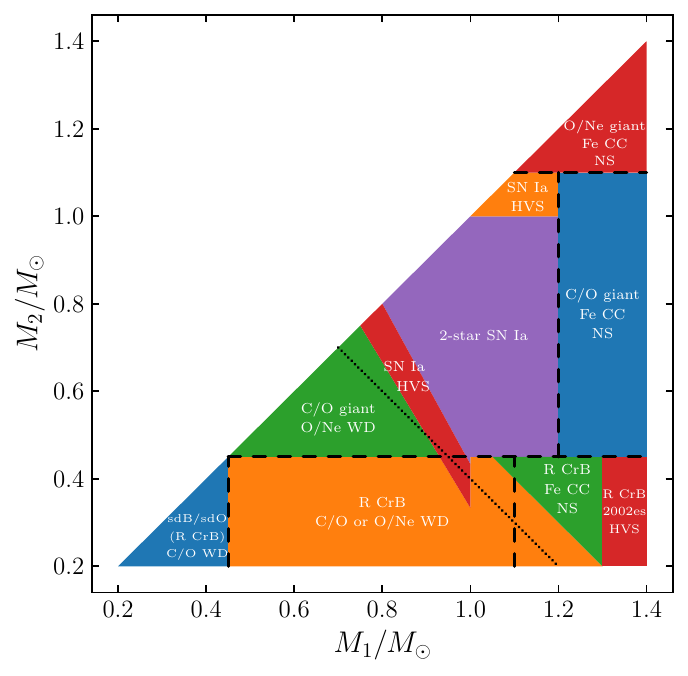}
	\caption{Same as Fig.\ \ref{fig:mvsm_noaddhe}, but assuming the primary WD has accreted additional He prior to the formation of the companion WD.  See text in Section \ref{sec:prevHe} for details.}
	\label{fig:mvsm_addhe}
\end{figure}

The work we have presented in this paper affords us a better  understanding of the evolution of interacting double WDs.  Motivated by these results, Figure \ref{fig:mvsm_noaddhe} shows a speculative set of possible outcomes of  double WD binaries, updating similar diagrams in \cite{dan14a} and \cite{shen15a}.  Dashed lines delineate boundaries between different core compositions, which we approximately take as $0.45 \msol$  between He and C/O cores and $1.1 \msol$ between C/O and O/Ne cores.  The dotted line demarcates a total mass of $1.4 \msol$.  We assume that all double WD binaries shown on this plot undergo unstable mass transfer, even if mass transfer initially occurs via a disk, due to the destabilizing effect of nova outbursts \citep{shen15a}.

The D$^6$ SN~Ia mechanism requires the accretion stream to strike the primary WD hard enough to ignite a He shell detonation and the primary WD to have a He layer of sufficient mass and density to sustain the detonation.  The latter requirement limits the primary to a mass $\lesssim 1.0 \msol$ \citep{shen24a}, although this mass can be increased via a previous phase of accretion, which we consider in the next section.

The criterion that the accretion stream is sufficiently violent constrains the minimum primary and secondary masses.  The gravitational potential, and thus the ram pressure of the incoming accretion stream, decrease with primary WD mass, and thus there should be a lower limit on the minimum primary mass. We set this lower mass somewhat arbitrarily at $0.75 \msol$, motivated by the dimmest SNe~Ia, the SN 1991bg-likes, which are well-matched by detonations of single $0.85 \msol$ WDs \citep{blon18a,shen18a,shen21a,shen21b} and likely by two-star $0.8+0.8 \msol$ explosions \citep{boos24a}.  

For a given primary mass, decreasing the secondary mass makes the incoming accretion stream's trajectory more grazing with respect to the primary's surface, thus decreasing the depth to which it penetrates and increasing the difficulty of igniting the He shell.  This implies a minimum companion mass that depends inversely on the primary mass.

If the secondary does not lose too much mass prior to the primary's explosion, it will undergo its own explosion when impacted by the ejecta, provided it is a C/O WD $\lesssim 1.0 \msol$ \citep{shen24a} or possibly a high-mass He WD \citep{papi15a,tani19a,boos24a}.  However, near the minimum companion mass limit, the mass transfer phase must last longer and become more violent \citep{dan11} before igniting the primary WD, and the companion will lose enough mass to avoid undergoing its own explosion, leaving behind a hypervelocity D$^6$ survivor.

The combination of the above considerations results in the purple region  in Figure \ref{fig:mvsm_noaddhe} containing two-star SNe~Ia (e.g., $0.8+0.9 \msol$) and the thin red region just below it (e.g., $0.45+0.95 \msol$)  giving rise to a single-star SN~Ia and a hypervelocity survivor (HVS).  We again emphasize that the boundaries are speculative and approximate but hopefully qualitatively informative.  Within this framework, D6-2 arose from the red region as a He WD that lost much of its mass before triggering the double detonation of its primary WD, which stripped almost all of the rest of the secondary's mass.  We also see that the same process should occur for low-mass C/O WDs.  These may result in survivors like D6-1 and D6-3, but if they lose as much mass as D6-2 has, we have seen in Section \ref{sec:d6} that they will be even redder and may still be awaiting discovery.

As discussed previously, high-mass primary C/O WDs do not have sufficient He envelopes at birth but may still explode via direct C ignition, but this necessitates very violent mass transfer.  This requires the near-disruption of a more massive secondary, and so the orange region to the right of the purple region (e.g., $0.9 + 1.05 \msol$) has a higher minimum secondary mass constraint. The large amount of unburned C/O may cause the primary's explosion to appear as a SN~2003fg-like event.  Furthermore, the copious mass loss from the companion means that it does not undergo its own explosion and instead survives as a low-mass hypervelocity star.

We assume that for binaries that do not explode during dynamical mass transfer, the less massive WD eventually becomes a spherically symmetric envelope surrounding the more massive WD that contracts until burning  is ignited at its base \citep{shen12,schw12,schw21a}.  This is the expected outcome for He + He WDs (lower left blue region, e.g., $0.3+0.4 \msol$; \citealt{schw18a}), He + low-mass C/O WDs (bottom orange region, e.g., $0.4+0.6 \msol$; \citealt{schw19b}), and low-mass C/O + C/O WDs (central green region, e.g., $0.6+0.6 \msol$; \citealt{schw21a}).\footnote{But c.f.\  \cite{pakm21a}, \cite{zena23a}, \cite{mora24a}, and \cite{glan24a} for simulations in which explosions are found in low-mass binaries.}  The nuclear burning proceeds as core- and/or shell-burning, depending on the core and shell compositions.  If a phase of shell-burning occurs, the merger remnant will appear as a giant star, possibly undergoing dust extinction events as in the case of He-burning R CrB stars.  The final fate of these binaries will be C/O or O/Ne WDs, likely bearing unique observational signatures such as fast rotation and strong magnetic fields  (although see \citealt{pakm24a}).  

Mass loss presumably plays an important role for higher-mass merger remnants.  We assume $\sim 10\%$ of the merger remnant's mass is lost during shell-burning, and thus relatively quiescent evolution in Figure \ref{fig:mvsm_noaddhe} extends to total masses of $1.5 \msol$.  Above this mass limit, the merging process may still proceed without an explosion, and the resulting merger remnant will still evolve similarly to the sequence of events outlined above, with shell-burning and/or core-burning phases, but the subsequent evolution does not lead to a surviving WD because the degenerate core reaches the Chandrasekhar mass.

In a C/O + O/Ne merger (right blue region, e.g., $0.8 + 1.3 \msol$; \citealt{wu23b}), the C/O WD secondary becomes an extended, C-burning envelope surrounding an O/Ne core that was the former primary WD.\footnote{The merger itself may result in a C detonation that partially burns the disrupted secondary, accompanied by a faint, rapidly evolving transient \citep{kash18a}, but little mass is ejected, and the remnant that is formed should evolve similarly to a remnant without a merger-induced detonation.}  The hot O/Ne ash that builds up underneath the C-burning layer eventually ignites and gives rise to a burning wave that propagates to the center of the core.  The subsequent evolution is similar to that of the cores of massive stars: successive phases of Ne-burning, O-burning, and Si-burning that convert the O/Ne core into an Fe core that eventually reaches the Chandrasekhar mass.

An O/Ne + O/Ne merger (upper right red region, e.g., $1.2+1.3 \msol$; \citealt{wu23a}) should evolve similarly to a C/O + O/Ne merger, but without a C shell-burning phase.  The outcome will be similar for both cases, though: the creation of an Fe core that grows to the Chandrasekhar mass.  This will result in an Fe core-collapse (Fe CC) SN that interacts with a relatively low-mass extended H- and He-deficient shell, appearing as a SN Icn \citep{fras21a,galy22a} or possibly a calcium-strong transient \citep{pere10,taub17a,jaco20a}, especially if the production of these otherwise rare binaries is dynamically enhanced in globular clusters \citep{shen19a}, which would explain their large observed galactocentric distances.  The resulting  NS may be rapidly rotating and highly magnetized, perhaps even giving rise to fast radio bursts \citep{krem21a,lu22a}.

For a He + relatively low-mass O/Ne merger (lower right green region, e.g., $0.4 + 1.2 \msol$; \citealt{broo17a}), shell-burning produces hot, degenerate ash, which will ignite and  yield a burning wave that  transforms the O/Ne core into an Fe core, eventually leading to a similar series of events as described above: an  Fe CC SN that interacts with an extended shell, possibly appearing as a calcium-strong transient, and the formation of a NS.

In a He + higher-mass O/Ne merger (lower right red region e.g., $0.4 + 1.3 \msol$), the O/Ne core will reach the Chandrasekhar mass prior to the ignition of  the He-burning ash and undergo an O deflagration.  Most previous studies have found that electron captures on the deflagration ashes cause the core to collapse into a neutron star in a so-called accretion-induced collapse (AIC; e.g., \citealt{miya80,nomo84b}).  However, recent work suggests the possibility that the O deflagration successfully propagates and ejects a portion of the star, leaving behind a kicked, hypervelocity remnant \citep{jone14a,jone16a,jone19a}.  Such hypervelocity survivors would be O/Ne-rich, which fits the atmospheric compositions of the LP $40-365$ class of hypervelocity stars \citep{venn17a,radd18b,radd19a,elba23a}.

An O deflagration is predicted to yield a relatively small amount of $^{56}$Ni \citep{jone16a}, but the collision of this ejecta with the pre-existing extended shell surrounding the O/Ne core will result in early-time shock interaction signatures \citep{broo17a}.  Subluminous SNe with early-time shock interaction match the characteristics of the SN~2002es-like class \citep{gane12,cao15a}.  He + higher-mass O/Ne WD binaries are rare in the field but may be overproduced dynamically in globular clusters \citep{shen19a}, thus explaining the  very large galactocentric distances observed for SN~2002es-like transients \citep{taub17a}.  Thus, we speculate that He + higher-mass O/Ne mergers  yield a shell He-burning phase, followed by a SN~2002es-like transient and the ejection of a hypervelocity remnant belonging to the LP $40-365$ class of hypervelocity stars.


\subsection{Outcomes for double WD  binaries with a previous phase of stable He mass transfer}
\label{sec:prevHe}

In Figure \ref{fig:mvsm_addhe}, we show a similar speculative set of interacting double WD outcomes as in Figure \ref{fig:mvsm_noaddhe}, but this time under  the assumption that the primary C/O WD undergoes a phase of thermal-timescale, dynamically stable He mass transfer when the companion is a non-degenerate He-burning star.  This phase of accretion can grow the mass of the primary WD \citep{yoon03a,ruit13a,broo16a}, and thus, we move the primary WD's C/O-O/Ne boundary to a representative higher mass of $1.2 \msol$.  As the He-burning companion shrinks within its Roche lobe, the mass transfer rate decreases, and the primary WD ceases burning He stably.  A series of He novae may then occur.  Further accretion at a decreasing rate is insufficient to continue igniting He novae, but crucially, the He surface layer on the primary WD will be left more massive than when it was born.

\cite{shen24a} found that C/O WDs $\gtrsim 1.0 \msol$ are born with He layers that are too low in mass to undergo a shell detonation.  However, the additional He that is gained by this subset of primary WDs that accreted mass prior to the formation of the double WD systems allows them to  undergo double detonations during future unstable mass transfer from their companion WDs.  This possibility increases the parameter space for two-star SNe~Ia (purple region in Fig.\ \ref{fig:mvsm_addhe}).

Moreover, high-mass companion WDs $\gtrsim 1.0 \msol$ (upper right orange region in Fig.\ \ref{fig:mvsm_addhe}; e.g., $1.0 + 1.1 \msol$) may be able to ignite a double detonation in these primaries with added He whilst being unable to undergo their own double detonation when the primary's ejecta impacts them, thus yielding one more channel to produce a surviving hypervelocity companion WD.  These companions would not undergo significant mass loss and would only have their surface layers heated by the SN ejecta impact.  Thus, they may be well-represented by the models calculated by \cite{bhat25a}, perhaps matching the hottest  D$^6$ survivors.


\section{Conclusions}
\label{sec:conc}

In this paper, we have calculated the evolution of initially fully convective stars of various compositions that were former accreting O/Ne degenerate cores that approached the Chandrasekhar mass and were partially disrupted in an explosion or former He or C/O WD companions to exploding primary C/O WDs.  The following are our main conclusions:

\bi

\item{The luminosities, effective temperatures, and ages of the three coolest D$^6$ survivors, and perhaps some of the others, can be explained by the  evolution of initially fully convective remnants of C/O and He companion WDs that lost the majority of their mass in the process of igniting their primary WDs.  Heating from tidal dissipation and the impact from the primary WD's ejecta, combined with expansion from the significant amount of mass loss, enabled the companions to be converted into fully convective stars.}

\item{If the primary WD's explosion occurs during the secondary's final plunge, the survivor's ejection velocity may be significantly higher than its pre-SN orbital velocity.}

\item{The rate of events that leave behind a hypervelocity survivor like D6-2 is $\sim 2\%$ of the SN~Ia rate, within small number statistics.  Explosions that release survivors like D6-1, D6-3, and possibly J1235-3752 make up $\sim 0.2\%$ of SNe~Ia, in line with the rate of SN~2003fg-like SNe.}

\item{The hottest D$^6$ survivors are likely cases in which only the surface layers of the companion WD were significantly altered by the primary WD's explosion.}

\item{The LP 40-365 stars can also be  explained by the Kelvin-Helmholtz cooling evolution of initially fully convective, low-mass stars.  These stars are consistent with having been an O/Ne WD (or having possessed a degenerate O/Ne core) that approached the Chandrasekhar mass and underwent a successful O deflagration that ejected most of its mass asymmetrically, kicking the low-mass surviving remnant at a high velocity.  Depending on the presence or absence of an extended shell, the explosion may appear as a SN~Iax or as a SN~2002es-like transient.}

\ei

The work presented in this paper is only a  step towards a complete understanding of hypervelocity SN survivors.  For one, our initially fully convective models are at best an approximation of the actual initial conditions of these stars.  Future suites of multidimensional hydrodynamic simulations  will provide better initial compositions, which may be altered by the deposition of thermonuclear ash, and entropy profiles, enabling further evolutionary studies.  They will also quantify the ejection velocities of companions that are released during their final plunge, thus exceeding their pre-SN orbital velocities.  On the observational side, future spectral modeling of the cool D$^6$ stars will test several of our conclusions by quantifying their masses and compositions.  Searches in all-sky spectroscopic surveys and in \emph{Gaia} data, possibly to redder colors, will find new candidates and better constrain the rates of thermonuclear SNe that give rise to hypervelocity survivors.


\software{\texttt{matplotlib} \citep{hunt07a}, \mesa \citep{paxt11,paxt13,paxt15a,paxt18a,paxt19a}}

\begin{acknowledgments}

This work is dedicated to Tom Marsh.  We thank Evan Bauer, Aakash Bhat, Lars Bildsten, Sam Boos, Liang Dai, Kareem El-Badry, Wenbin Lu, Alison Miller, R\"{u}diger Pakmor, Eliot Quataert, Roberto Raddi, Dean Townsley, and Sunny Wong for helpful discussions, Josiah Schwab for assistance with opacity tables, and the referee for useful comments.  This work was supported by NASA through the Astrophysics Theory Program (80NSSC20K0544)  and by NASA/ESA Hubble Space Telescope programs \#15871, \#15918, and \#17441.  This research used the Savio computational cluster resource provided by the Berkeley Research Computing program at the University of California, Berkeley (supported by the UC Berkeley Chancellor, Vice Chancellor of Research, and Office of the CIO).

\end{acknowledgments}



\end{document}